\documentclass[twocolumn,aps,showpacs,prb,tightenlines,amsmath,amssymb,superscriptaddress]{revtex4}
\usepackage{graphicx}
\usepackage{amssymb}
\usepackage{dcolumn}
\usepackage{amsmath}
\usepackage{bm}

\usepackage{colordvi}

\begin{document}

\title{Electron spin relaxation in cubic GaN quantum dots}

\author{M. Q. Weng}
\affiliation{Hefei National Laboratory for Physical Sciences at
Microscale,
University of Science and Technology of China, Hefei,
Anhui, 230026, China}
\affiliation{Department of Physics,
University of Science and Technology of China, Hefei,
Anhui, 230026, China}
\author{Y. Y. Wang}
\affiliation{Department of Physics,
University of Science and Technology of China, Hefei,
Anhui, 230026, China}
\author{M. W. Wu}
\thanks{Author to whom correspondence should be addressed}
\email{mwwu@ustc.edu.cn.}
\affiliation{Hefei National Laboratory for Physical Sciences at
Microscale,
University of Science and Technology of China, Hefei,
Anhui, 230026, China}
\affiliation{Department of Physics,
University of Science and Technology of China, Hefei,
Anhui, 230026, China}
\altaffiliation{Mailing address}

\date{\today}

\begin{abstract}

  The spin relaxation time $T_{1}$ in zinc blende GaN quantum dot is
  investigated  for different magnetic field,
  well width and quantum dot diameter. The spin relaxation caused by 
  the two most important spin relaxation mechanisms in zinc blende
  semiconductor quantum dots, {i.e.} the electron-phonon
  scattering in conjunction with the Dresselhaus spin-orbit coupling
  and the second-order process of the hyperfine interaction combined
  with the electron-phonon scattering, are systematically studied. The
  relative importance of the two mechanisms are compared in detail
  under different conditions. It is found that due to the small spin
  orbit coupling in GaN, the spin relaxation caused by the
  second-order process of the hyperfine interaction combined with the
  electron-phonon scattering plays much more important role than it
  does in the quantum dot 
  with narrower band gap and larger spin-orbit coupling, such as GaAs
  and InAs.

\end{abstract}

\maketitle

\section{Introduction}

The wide-band-gap group III nitride semiconductor GaN has emerged as a
leading material for a variety of new devices,\cite{Gil,Grandjean}
ranging from the blue laser\cite{Nakamura,Nakamura2} to high-power
electronic devices,\cite{Morkoc} by 
utilizing its electronic and optical properties. 
Recently the magnetic properties of GaN-based nanostructures have also 
attracted much attention, due to the potential application
in spintronic device.\cite{Pearton} 
Understanding the carrier spin relaxation mechanism in GaN is of
great importance in the design and the realization of GaN-based spin
device. 
So far, much effort has been devoted to the experimental study of 
the spin relaxation in different GaN structures, including GaN
epilayers,\cite{Beschoten, Kuroda, Ishiguro, Tackeuchi} GaN 
quantum wells\cite{Nagahara,Nagahara2,Chen} and GaN quantum dots
(QDs).\cite{Lagarde}
Most of these works focus on the spin life time in the
hexagonal wurtzite GaN structures, which are easier to grow than
the cubic structures. However, the
spin-orbit coupling (SOC) in wurtzite GaN structure is much larger
than that of cubic
GaN due to the strong built-in electric field caused by the spontaneous and
piezoelectric polarizations.\cite{Fonoberov,Julier} The
electron/exciton spin life time of 
different wurtzite GaN nanostructures ranges from a few to a few hundred
picoseconds.
 While the exciton spin relaxation time is of nanoseconds
for in the cubic GaN epilayer\cite{Tackeuchi} and is even longer in
cubic GaN QD.\cite{Lagarde}
On the theoretical side, spin relaxation times of electron and hole in 
bulk cubic GaN are calculated and are found to be two or three orders of 
magnitude longer than those in GaAs.\cite{Krishnamurthy,Yu1}  
However, the electron spin properties in cubic GaN QDs are less well
understood and many questions, such as what the dominant spin
relaxation mechanism is, remain open. 
In this paper, we will systematically study the electron spin
relaxation in cubic GaN QD under different conditions.

There are many spin relaxation mechanisms in
QDs.\cite{Khaetskii,Woods,Cheng,Erlingsson,Jiang,Voss} In cubic  
semiconductor QDs, the most important two mechanisms are: 
(1) the electron-phonon scattering in conjunction with the SOC,
and (2) the second-order process of the hyperfine interaction combined
with the electron-phonon scattering.\cite{Erlingsson,Jiang,Voss} In GaAs
QD, it was shown that the first mechanism 
is the dominant spin relaxation mechanism for quite wide range of
parameters due to the large SOC.\cite{Jiang} 
Since the SOC in GaN is much smaller than that of GaAs, which of these
two mechanisms dominates spin relaxation need to be further
examined. 
 
We organize the paper as following. In Sec.~\ref{sec:model} we 
set up the model and give the Hamiltonian. The two most important
electron spin relaxation mechanisms are discussed and the formula of
the corresponding spin relaxation rates are presented. We then
calculate the spin relaxation rates 
of a QD embedded in a narrow quantum well
analytically using perturbation
theory in Sec.~\ref{sec:an_result}. We further present the exact spin
relaxation rates under different conditions by numerical method in
Sec.~\ref{sec:results} and summarize in Sec.~\ref{sec:conclusion}. 

\section{Model and Spin Relaxation Rate}
\label{sec:model}

We consider one electron spin in a single GaN QD embedded in a quantum
well with well width $a$. A magnetic field ${\mathbf B}$ is applied. 
The Hamiltonian of the system composed of the electron and the 
lattice is given by:
\begin{equation}
H_{T}=H_{e}+H_{L}+H_{eL}\ ,
\label{eq:HT}
\end{equation}
where $H_{e}$, $H_{L}$ and $H_{eL}$ are the Hamiltonians of the
electron, the lattice and their interaction respectively.
The electron Hamiltonian  $H_{e}$ can be written as: 
\begin{eqnarray}
  H_{e} &=& H_{0}+H_{SO} \nonumber \\ 
  &=& [\frac{\mathbf{P}^{2}}{2m^\ast}
  +V_c(x,y)+V_z(z) + H_Z] + H_{SO}\ ,
\label{eq:He}
\end{eqnarray}
where $H_{0}$ is the Hamiltonian 
without the SOC, $m^{\ast}$ is the
electron effective mass and $\mathbf{P}=-i \hbar \mathbf{\nabla} +
\frac{e}{c} \bf{A}$ is the 
kinetic momentum with 
${\mathbf A}={\mathbf B}\times {\mathbf
    r}$.  $V_z(z)$ is the quantum well confinement. In this paper, it
  is assumed to be a hard wall confinement with width $a$. $V_{c}(x,y)
  =\frac{1}{2} m^{\ast} 
  \omega_{0}^{2} (x^2+y^2)$ is the in-plane confinement of QD with
  diameter  
$d_{0}=\sqrt{\hbar\pi/m^{\ast} \omega_{0}}$.
$H_Z=\frac{1}{2}g\mu_B {\mathbf B}\cdot\mbox{\boldmath
  $\sigma$\unboldmath}$ is the Zeeman
energy with $g$, $\mu_{B}$ and $\mbox{\boldmath
  $\sigma$\unboldmath}$ being the $g$-factor
of electron, Bohr magneton and Pauli matrix, respectively. 
$H_{so}$ is the Hamiltonian of the SOC.
In cubic GaN the dominant SOC term is Dresselhaus term,\cite{Dresselhaus}
which reads\cite{rashba_2003} 
\begin{eqnarray}
&&H_{so}=\frac{1}{\hbar^{3}}\gamma_{0}
[(P_yP_{x}P_y-P_{z}P_xP_z)\sigma_{x}+ 
(P_zP_{y}P_{z} \nonumber\\ && 
-P_{x}P_yP_x)\sigma_{y}
+ (P_xP_{z}P_{x}-P_{y}P_zP_y)\sigma_{z}]\ , 
\end{eqnarray}
with $\gamma_{0}$
being the Dresselhaus coefficient.
The eigen wave function $|\ell\rangle$ and the eigen
energy $\varepsilon_{\ell}$ ($\ell=1,2,\cdots$) of $H_{e}$ can be
obtained from the perturbation theory or 
from the exact diagonalization method,\cite{Cheng} using the
eigenstates of $H_{0}$ as basis. 
The Hamiltonian of the lattice is consisted of two parts:
$H_{L}=H_{ph}+ H_{nuclei}$. $H_{ph}=\sum_{{\mathbf q}\eta} \hbar
\omega_{{\mathbf q}\eta}a_{{\mathbf q}\eta}^{\dagger} a_{{\mathbf
    q}\eta}$ represents the Hamiltonian of the phonons with
$\omega_{{\mathbf q}\eta}$ standing for the phonon energy spectrum of
branch $\eta$ and momentum ${\mathbf q}$ and
$a^{\dagger}_{\mathbf{q}\eta}$/$a_{\mathbf{q}\eta}$ 
being the corresponding phonon creation/annihilation operator.
$H_{nuclei}=\sum_{j} \gamma_{I} {\mathbf B}\cdot {\mathbf I}_{j}$
is the Zeeman term of the the lattice nuclear spins in the
external magnetic field with $\gamma_{I}$ 
and ${\mathbf I}_{j}$ denoting the gyro-magnetic ratio and spin of
$j$-th nucleus respectively.  
The interaction between the electron and the lattice also
has two parts: $H_{eL}=H_{ep}+H_{eI}$. $H_{ep}$ is the
electron-phonon scattering and is given by
$H_{ep}=\sum_{{\mathbf q}\eta} M_{{\mathbf q}\eta} 
(a_{{\mathbf q}\eta} + a^{\dagger}_{-{\mathbf q}    \eta})
e^{i {\mathbf q} \cdot {\mathbf r}}$,
where $M_{{\mathbf q}\eta}$ is the matrix element of the electron-phonon
interaction. 
$|M_{{\mathbf q}sl}|^{2} = \hbar \Xi^{2}q/2 \rho v_{sl}$ for the
electron-phonon
coupling due to the deformation potential. For the
piezoelectric coupling,
$|M_{{\mathbf q}  pl}|^{2}= (32\hbar \pi^{2}
e^{2}e_{14}^{2}/\kappa^{2} \rho
v_{sl})[(3q_{x}q_{y}q_{z})^{2}/q^{7}]$ for the longitudinal phonon
mode and   $\sum_{j=1,2}|M_{{\mathbf q}pt_j}|^{2} =[32\hbar \pi^{2}
e^{2}e_{14}^{2}/(\kappa^{2} \rho
v_{st}q^{5})][q_{x}^{2}q_{y}^{2}+q_{y}^{2}q_{z}^{2} +
q_{z}^{2}q_{x}^{2} -(3q_{x}q_{y}q_{z})^{2}/q^{2}]$ for the two transverse modes.
Here $\Xi$
stands for the acoustic deformation potential;
$\rho$ is the GaAs volume density; 
$e_{14}$
is the piezoelectric constant and $\kappa$ denotes
the static dielectric constant. The acoustic phonon spectra
$\omega_{{\mathbf q}l} = v_{sl} q$ for the longitudinal mode and
$\omega_{{\mathbf q}t} = v_{st}q$ for the transverse modes with
$v_{sl}$ and $v_{st}$ representing the corresponding sound 
velocities. 
$H_{eI}$ is the electron-nucleus hyperfine
interaction $H_{eI}$, which can be written as
$H_{eI}=\sum_{j}Av_{0}{\mathbf S}\cdot {\mathbf I}_{j} \delta ({\mathbf
r}- {\mathbf R}_{j})$,
where $v_{0}$ is the volume of the unit cell of the lattice, ${\mathbf S}$ is the spin of the electron,  ${\mathbf r}$ and ${\mathbf
  R}_{j}$ are the  position of
the electron  and the $j$-th nucleus, respectively. 
$A$ stands for the hyperfine interaction coupling
constant.

In the Hamiltonian [Eq.~(\ref{eq:HT})], we only include the terms 
related to the two dominant spin relaxation mechanisms. One is the
electron-phonon scattering in conjunction with the Dresselhaus
SOC. The SOC mixes the spin-up and -down states to form the majority
spin-up and -down states. The direct coupling to the phonon causes the
transition between the majority spin-up and -down states and results
in the spin relaxation. The transfer matrix element is 
$M_{{\mathbf q}\eta}$. This spin mechanism is referred as 
``Mechanism I'' hereafter. 
The other is the second-order process of the hyperfine interaction
combined with the electron-phonon interaction in which not only
the SOC mixes the spin-up and spin-down states, but also the nuclei flip
the electron spin. 
As the phonon compensates the energy
difference, this mechanism also leads to spin relaxation. 
In the following, it is called ``Mechanism II''. 
The transfer matrix between states $|\ell_1\rangle$ and
$|\ell_2\rangle$ of Mechanism II 
can be written as
\begin{eqnarray}
V_{eI-ph} &=&  | \ell_{2}
\rangle\biggl[ \sum_{m\not=\ell_1} \frac{ \langle
\ell_{2} | H_{ep} |m \rangle \langle m | {H_{eI}
} |\ell_{1}
\rangle }{ \varepsilon_{\ell_{1}} - \varepsilon_{m}} \nonumber \\
&&+  \sum_{m\not=\ell_2} \frac{ \langle
\ell_{2} |{H_{eI} 
} |m \rangle \langle m |H_{ep}|\ell_{1}
\rangle }{\varepsilon_{\ell_{2}} -\varepsilon_{m}}\biggr]\langle \ell_{1}|
\nonumber \\
&=& \sum_{{\mathbf q} \eta} {\cal M}_{{\mathbf q}\eta} (a_{{\mathbf
    q}\eta} + a^{\dagger}_{-{\mathbf q}    \eta})\ ,
\label{eq:HeI}
\end{eqnarray}
with
\begin{eqnarray}
{\cal M}_{{\mathbf q}\eta}&=&| \ell_{2}
\rangle \biggl[ \sum_{m\not=\ell_1} \frac{ \langle
\ell_{2} | M_{{\mathbf q}\eta}e^{i {\mathbf q}\cdot {\mathbf r}} |m
\rangle \langle m | {H_{eI}
} |\ell_{1}
\rangle }{ \varepsilon_{\ell_{1}} - \varepsilon_{m}} \nonumber \\
&&+  \sum_{m\not=\ell_2} \frac{ \langle
\ell_{2} |{H_{eI} 
} |m \rangle \langle m | M_{{\mathbf q}\eta}e^{i {\mathbf q}\cdot
  {\mathbf r}}|\ell_{1} 
\rangle }{\varepsilon_{\ell_{2}} -\varepsilon_{m}}\biggr]\langle
\ell_{1}|\ ,
\label{eq:MeI}
\end{eqnarray}
where the summation of $|m\rangle$ runs over all possible 
intermediate states.

To calculate the spin relaxation time, one can use the perturbative
approach based on the calculation of the transition rates from Fermi's
golden rule.\cite{Khaetskii,Woods,Cheng,Golovach,Voss} 
Non-perturbative calculation using equation of motion method has also
been proposed to study the spin relaxation of the system with
large SOC at high temperature regime.\cite{Jiang,Voss} For the system with
weak SOC at low temperature regime, these two approaches produce the
same results. In the cubic GaN QD, since the SOC is pretty weak,\cite{Fu} 
the perturbative approach gives sufficient accurate spin relaxation
rate and is therefore adopted in the present work.

Using the Fermi's golden rule, one can obtain the spin relaxation rate
as:\cite{Jiang}  
\begin{equation}
T_{1}^{-1}=\sum_{if}( f_{i+}\Gamma_{i+\to f-}+ f_{i-}\Gamma_{i-\to f+})\ .
\end{equation}
Here $f_{i\pm}$ is the Maxwell distribution
since we study the spin relaxation of single electron confined in
  the QD.
``$+/-$'' stand for the states with the majority up/down-spin. 
The scattering rate $\Gamma_{i\to f}$ reads:
\begin{eqnarray}
  &&\hspace{-0.1cm}
  \Gamma_{i\to f}=\frac{2\pi}{\hbar}\sum_{{\mathbf q}\eta} |\langle
  f|{\cal X}_{{\mathbf q}\eta} |i\rangle|^{2} 
  \bigl[n_{{\mathbf q}\eta} \delta
  (\varepsilon_{f}-\varepsilon_{i}-\hbar \omega_{{\mathbf q}\eta})
  \nonumber \\
  &&\hspace{1.2cm}
  \mbox{}+(n_{{\mathbf q}\eta}+1)\delta
  (\varepsilon_{f}-\varepsilon_{i}+\hbar 
  \omega_{{\mathbf q}\eta}) \bigr]\ ,
\end{eqnarray}
where ${\cal X}_{{\mathbf q}\eta}=M_{{\mathbf q}\eta} e^{i{\mathbf
    q}\cdot {\mathbf r}}$ and 
${\cal X}_{{\mathbf q}\eta}={\cal M}_{{\mathbf q}\eta}$
for Mechanism I and II, 
respectively.
$n_{{\mathbf q}\eta}$ is the Bose distribution function for phonons. 


\section{Analytical Results}
\label{sec:an_result}

Before presenting the full exact diagonalization result, let
us first look at the analytical result of the spin relaxation
  rate of a QD embedded in a narrow quantum well
by perturbatively solving the electron
Hamiltonian [Eq.~(\ref{eq:He})] to the second order of the SOC.

Due to the symmetry of the QD in the $x$-$y$ plane, ${\mathbf B}$ can
be assumed to be $(B_{\|},0,B_{\perp})$ with $B_{\|}=B \sin \theta$
and $B_{\bot}=B\cos \theta$ being the components along the $x$- and
$z$-axis, and $\theta$ representing the angle between the magnetic
field direction and the $z$-axis.  
The eigenstate of the electron
Hamiltonian without the SOC ($H_{0}$) $|n_z nl\sigma\rangle$ is
characterized by the quantum number of quantum well confinement, 
radial, angular and spin
freedoms $n_z\ (=1,2,\cdots)$,
$n\ (=0,1,\cdots)$, $l\ (=0,\pm 1, \cdots)$ and $\sigma\
(=\pm 1)$ respectively, whose energy is $E_{n_z nl\sigma}=
{n_z^2\hbar^2\pi^2\over 2m^{\ast}a^2}+
(2n+|l|+1)\hbar
\Omega +l\hbar\omega_{B_{\bot}}+ \sigma E_{B}$, with
$\Omega=\sqrt{\omega_{0}^{2} + \omega_{B_{\bot}}^{2}}$,
$\omega_{B}=eB_{\bot}/(2m^{\ast})$ and
$E_{B}=\frac{1}{2}g\mu_{B}B$. 
  In the narrow quantum well ($d_0\gg a$), the distance of different
  $n_z$ states is so large that only the lowest $n_z$ state is
  relevant. Under this approximation, the spin orbit coupling can be
  expressed as 
$H_{so}= \frac{1}{\hbar}\gamma_{0}(\pi/a)^2(-P_{x}\sigma_{x}+P_{y}\sigma_{y})$.
Up to the first order perturbation, 
the lowest two eigenstates of the electron with the SOC are  
\begin{widetext}
\begin{eqnarray}
  \label{eq:state}
 |\ell_{1}\rangle&=&|001\rangle+{\cal B}_{011,001}^{1,+} |011\rangle - {\cal
   B}_{01-1,001}^{3,+}|01-1\rangle+ {\cal B}_{0-11,001}^{1,-}
 |0-11\rangle +{\cal B}_{0-1-1,001}^{2,-} |0-1-1\rangle \ , \nonumber \\
 |\ell_{2}\rangle&=&|00-1\rangle+{\cal B}_{011,00-1}^{2,+}|011\rangle-{\cal
   B}_{01-1,00-1}^{1,+}|01-1\rangle - {\cal B}_{0-11,00-1}^{3,-}|0-11\rangle-{\cal
   B}_{0-1-1,00-1}^{1,-}|0-1-1\rangle\ , 
\end{eqnarray}
where ${\cal B}_{nl\sigma,n^{\prime}l^{\prime}\sigma^{\prime}}^{1,\pm}=i\frac{1}{2}\hbar
\gamma_{c}\alpha (1\pm \omega_{B_{\bot}}/\Omega)\sin
\theta/(E_{nl\sigma}-E_{n^{\prime} l^{\prime}\sigma^{\prime}})$, ${\cal B}_{nl\sigma,n^{\prime}l^{\prime}\sigma^{\prime}}^{2,\pm}=i\frac{1}{2}\hbar
\gamma_{c}\alpha (1\pm \omega_{B_{\bot}}/\Omega)(1+\cos
\theta)/(E_{nl\sigma}-E_{n^{\prime} l^{\prime}\sigma^{\prime}})$ and ${\cal B}_{nl\sigma,n^{\prime}l^{\prime}\sigma^{\prime}}^{3,\pm}=i\frac{1}{2}\hbar
\gamma_{c}\alpha (1\pm \omega_{B_{\bot}}/\Omega)(1-\cos
\theta)/(E_{nl\sigma}-E_{n^{\prime} l^{\prime}\sigma^{\prime}})$.
The corresponding eigen energies of these states read
\begin{eqnarray}
  \label{eq:eng}
\varepsilon_{1}&=&E_{001}+|{\cal B}_{011,001}^{1,+}|^{2}
(E_{011}-E_{001})-|{\cal B}_{01-1,001}^{3,+}|^{2} (E_{01-1}-E_{001})
+|{\cal B}_{0-11,001}^{1,-}|^{2} (E_{0-11}-E_{001})\nonumber \\
&&+|{\cal B}_{0-1-1,001}^{2,-}|^{2} (E_{0-1-1}-E_{001})\ ,
\nonumber \\
\varepsilon_{2}&=&E_{00-1}+|{\cal B}_{011,00-1}^{2,+}|^{2}
(E_{011}-E_{00-1}) - |{\cal B}_{01-1,00-1}^{1,+}|^{2}
(E_{01-1}-E_{00-1}) -|{\cal B}_{0-11,00-1}^{3,-}|^{2}
(E_{0-11}-E_{00-1})\nonumber \\
&& -|{\cal B}_{0-1-1,00-1}^{1,-}|^{2}
(E_{0-1-1}-E_{00-1})\ .
\end{eqnarray}
\end{widetext}
It is noted that in the above equations, we have included the
second-order correction of the SOC on the energy, which is crucial to 
the study of the spin relaxation using perturbation method
as pointed out by Cheng {\em et al.}\cite{Cheng}
It is also noted that the state index $n_z$ is dropped in the above
equations since it is always $1$ in the narrow quantum well
approximation. 
$|\ell_1\rangle$ and $|\ell_2\rangle$ are the lowest majority spin-up and
spin-down states respectively. At the low temperature regime, the
electron mainly distributes on these two states. Therefore, $T_{1}$
basically equals the spin relaxation time between these two states.  
It is further noted that at low temperature regime, the main electron-phonon
scattering comes from the electron coupling to  
the transverse phonon via piezoelectric field. 
With these approximations, the spin relaxation rate due to 
Mechanism I is given by
\begin{widetext}
\begin{eqnarray}
  \Gamma_{1} &= & c (2n_{{\mathbf q}}+1)\alpha q 
\int_{0}^{\pi/2} d \theta^{\prime} \sin^{3} \theta^{\prime} (\sin^{4}
\theta^{\prime} + 8\cos^{4} \theta^{\prime} )  e^{-q^{2} \sin^{2} \theta^{\prime}/2}
 I^{2}(\frac{1}{2}q a\alpha \cos \theta^{\prime}) [2{\cal
   P}_{1}^{2} \nonumber \\
&&+ ({\cal P}_{2}^{2}+{\cal P}_{3}^{2}-2{\cal P}_{1}^{2}) \frac{1}{4}
 q^{2} \sin^{2} \theta^{\prime} +({\cal P}_{4}^{2}+{\cal
   P}_{5}^{2}+
   2{\cal P}_{1}^{2}) \frac{1}{16}q^{4} \sin^{4} \theta^{\prime}]\ ,
\label{phonon}
\end{eqnarray}
\end{widetext}
where $c=\pi e^{2}e_{14}^{2}/(\hbar D v_{st}^{2} \kappa^{2})$,
$q=\Delta E/(\hbar v_{st}\alpha)$ with $\Delta
E=|\varepsilon_{2}-\varepsilon_{1}|$ and
$\alpha=\sqrt{m^{\ast}\Omega/\hbar}$.
$I(x)= \pi^{2} \sin(x)/[x(\pi-x)(\pi+x)]$ denotes the form factor
along $z$-direction due to the quantum well confinement. 
In the above equation, ${\cal P}_{1}={\cal A}_{1}+{\cal A}_{2}-{\cal A}_{3}-{\cal
  A}_{4}$ with  ${\cal
  A}_{1}=|{\cal B}_{011,001}^{1,+}{\cal B}_{011,00-1}^{2,+}|$, ${\cal
  A}_{2}= |{\cal B}_{01-1,001}^{3,+} {\cal B}_{01-1,00-1}^{1,+}|$,
${\cal A}_{3}=|{\cal B}_{0-11,001}^{1,-}{\cal B}_{0-11,00-1}^{3,-}|$ and
${\cal A}_{4}=|{\cal B}_{0-1-1,001}^{2,-}{\cal
  B}_{0-1-1,00-1}^{1,-}|$; 
 ${\cal P}_{2}=-|{\cal B}_{011,00-1}^{2,+}|+|{\cal
  B}_{0-11,00-1}^{3,-}|-| {\cal B}_{01-1,001}^{3,+}|+|{\cal
  B}_{0-1-1,001}^{2,-}|$; ${\cal P}_{3}=-|{\cal B}_{011,00-1}^{2,+}|-|{\cal
  B}_{0-11,00-1}^{3,-}|+| {\cal B}_{01-1,001}^{3,+}|+|{\cal
  B}_{0-1-1,001}^{2,-}|$, ${\cal P}_{4}={\cal C}_{1}-{\cal
  C}_{2}+{\cal C}_{3}-{\cal C}_{4}$, 
${\cal P}_{5}={\cal C}_{1}-{\cal
  C}_{2}-{\cal C}_{3}+{\cal C}_{4}$ with ${\cal C}_{1}=|{\cal
  B}_{011,001}^{1,-}{\cal B}_{0-11,00-1}^{3,+}|$, ${\cal C}_{2}=|{\cal
  B}_{01-1,001}^{1,-}{\cal B}_{0-1-1,00-1}^{3,+}|$, ${\cal C}_{3}=|{\cal
  B}_{0-11,001}^{1,-}{\cal B}_{011,00-1}^{2,+}|$ and ${\cal C}_{4}=|{\cal
  B}_{0-1-1,001}^{1,-}{\cal B}_{01-1,00-1}^{2,+}|$.
Using the material parameters of GaN QD and in consideration of the
relative small magnetic field,
one can write down the
spin relaxation rate due to this mechanism at zero
temperature for relative small dot:
\begin{eqnarray}
\Gamma_1&\propto& a^{-4}d_{0}^{8} B^{5} (1+\cos^{2}
\theta)\ ,
\label{gamma1}
\end{eqnarray}
which indicates that for fixed magnetic field magnitude, the spin
relaxation under the perpendicular magnetic field is
two times of that under the parallel magnetic filed.
It should be noted that for $\theta=0$ case (the magnetic field is
along the $z$-axis), to the leading term, the magnetic field 
dependence of $\Gamma_{1}$ obtained here is in
accordance with that obtained in Refs.~\onlinecite{Khaetskii} and 
\onlinecite{Woods}.
By assuming that the nuclei spins are independent to each other and are
  in equilibrium state, 
the spin relaxation between $|\ell_1\rangle$ and $|\ell_2\rangle$
induced by Mechanism II, with the mediation of the lowest
  available state, 
can be written as  
\begin{eqnarray}
\Gamma_{2}&=&(\frac{A}{\varepsilon_{2}-\varepsilon_{3}})^{2} I(I+1)
v_{0}\alpha^{3} a^{-1} c (2n_{{\mathbf q}}+1) q^{3}\nonumber\\ &&
\times \int_{0}^{\pi/2} d \theta^{\prime} \sin^{5} \theta^{\prime} 
(\sin^{4} \theta^{\prime} + 8\cos^{4}
\theta^{\prime} ) \\ &&
\times e^{-q^{2} \sin^{2} \theta^{\prime}/2}
 I^{2}(\frac{1}{2}q a\alpha \cos \theta^{\prime})\ , \nonumber
\label{second-order}
\end{eqnarray}
which at zero temperature gives
\begin{equation}
\Gamma_{2} \propto c_{3} a^{-1} d_0^4B^3 \ .
\label{gamma2}
\end{equation}
The ratio of the spin relaxations due to these two mechanisms
is therefore 
\begin{eqnarray}
  \label{eq:rate_ratio}
  \Gamma_{1}/\Gamma_{2}
\propto a^{-3}d_0^4 B^2\ ,
\end{eqnarray}
which gives a guideline to determine which mechanism is more
important at different conditions. It is therefore expected that 
Mechanism II is more 
important for smaller QD embed in wider quantum well under 
weaker magnetic field. 

\section{Numerical Results}
\label{sec:results}

\begin{table}[htb]
\caption{Parameters used in the calculation.}
\hspace{2cm}
\begin{tabular}{p{1cm} p{3.5cm} p{1cm} l }
\hline \hline
$\rho$ &  $6.095 \times 10^{3}$\ kg/m$^{3}$ & $\kappa$ & 8.5 \\
$v_{st}$  & $2.68 \times 10^{3}$\ m/s & $g$ & $2.06$ \\
$v_{sl}$ & $6.56 \times 10^{3}$\ m/s & $\Xi$ & $8.3$\ eV \\
$e_{14}$ & $4.3 \times 10^{9}$\ V/m & $m^{\ast}$ & $0.15 m_{0}$ \\
$A$    & $45$\ $\mu$eV    & $I$ & $\frac{3}{2}$  \\
\hline \hline
\end{tabular}
\label{table1}
\end{table}

The perturbation method gives qualitative results for us to understand
the overall behavior of spin relaxation in GaN under different
conditions. However, in the perturbation calculation, states with
higher energy are dropped to get a manageable analytical result.  
It should be noted that, for the spin relaxation caused by 
Mechanism II, 
the contributions of higher
intermediate states and the lowest one are of the same order in regard
to hyperfine interaction strength. 
  Moreover, for QD embedded in wider quantum wells, contribution of
  the higher $n_z$ states to the spin orbit coupling can
  not be neglected. It is expected that for $d_0\leq a$, the
  spin relaxation due to Mechanism I can be different from the
  perturbative results. 
It is therefore necessary to
check the accuracy of the perturbative calculation by comparing to
the exact diagonalization with sufficient basis functions included. 

\begin{figure}[bth]
\centering
  \includegraphics[width=1.\columnwidth]{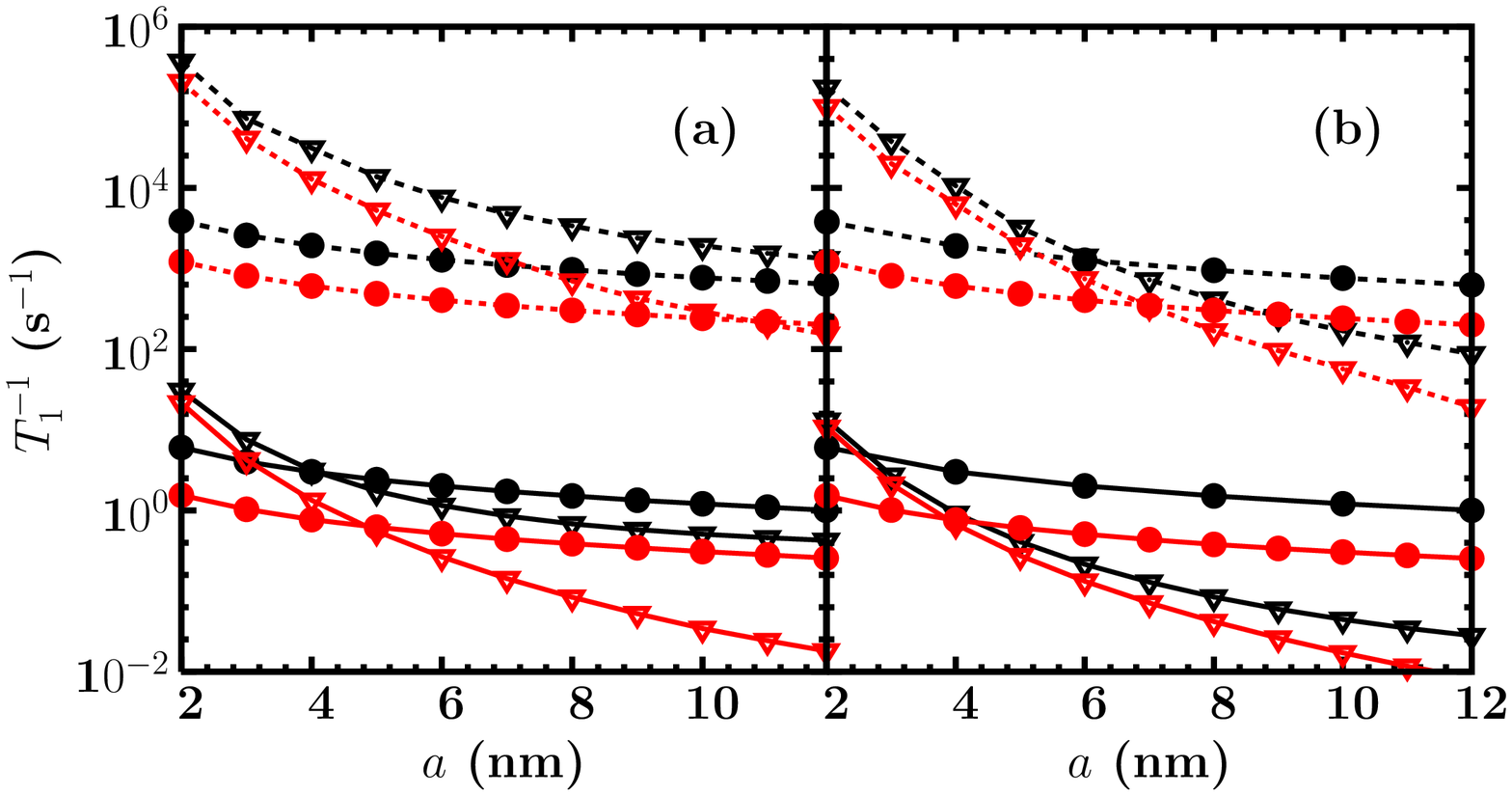}
  \caption{(Color online) Spin relaxation rate {\em vs.}
    the well width $a$
  in the presence of (a) perpendicular and (b) parallel magnetic fields
with  $B=0.5$\ T (solid curves) and $B=5$\ T (dotted
  curves). In the calculation,
    $d_{0}=10$\ nm. Black (dark) curves---exact diagonalization results;
Red (light) curves---perturbation results. 
  Curves with $\bigtriangledown$---$T_{1}^{-1}$ induced by
    the electron-phonon scattering in conjunction with the SOC; Curves
    with $\bullet$---$T_{1}^{-1}$ induced by the second-order process of the
    hyperfine interaction combined with the electron-phonon scattering.}
  \label{fig:width}
\end{figure}

In Fig.~\ref{fig:width}, we present the spin relaxation rates as
functions of well width in GaN QD under different conditions obtained
by the exact diagonalization and perturbation. The material parameters
of GaN are listed in
Table~\ref{table1}.\cite{Madelung,Vurgaftman,Albrecht,Krummheuer} The
Dresselhaus coefficient  $\gamma_0$ is chosen to be
$0.51$~\AA$^3\cdot$eV according to the latest calculation.\cite{Fu} It
is seen that the perturbation results describe the qualitative 
behavior of the spin relaxation 
pretty well. 
For the cases we study here, 
the spin relaxation caused by the electron-phonon scattering in
conjunction with the SOC from the perturbation is very close to the
exact diagonalization result in narrow quantum well. When the 
well width becomes larger, the difference between perturbative and
exact diagonalization result also grows as contribution of the higher $n_z$
states becomes more and more important. 
On the other hand, for the spin
relaxation caused by the hyperfine interaction together with
electron-phonon interaction,
the difference between perturbative and exact diagonalization results
almost doest not change with the well width. For this mechanism, 
the relaxation rate from
exact diagonalization method is always about one order of magnitude
higher than that obtained from the perturbation calculation for the
cases we study.
This indicates the contribution of the higher states are important to
the quantitative calculation of the spin relaxation. 
In the following, we only present the results of exact
diagonalization unless otherwise  specified. 
We now focus on how the spin relaxation induced by the two
mechanisms change with $a$. 
It is seen that the spin relaxations induced by the two
mechanisms both decrease with $a$. The spin relaxation due to 
Mechanism I
decreases much faster than that due to Mechanism II.
As a result, Mechanism II 
becomes more and more
important when the quantum well width increases. 
This can be understood from the perturbation result. As one
can see from Eqs.~(\ref{gamma1}) and (\ref{gamma2}) that the relaxation
rate $\Gamma_1$ decreases with $a$ as $a^{-4}$, while $\Gamma_2$ is
proportional to $a^{-1}$. We then pay attention to the relative
  importance of these two mechanisms.
For the vertical-magnetic-field case in
Fig.~\ref{fig:width}(a), for $B=5$~T, 
spin relaxation due to Mechanism I is always the dominant spin
  relaxation mechanism. 
 When $B$ decreases to $0.5$\ T, 
Mechanism II 
almost dominates the spin relaxation 
except at very small well width ($a<4$\ nm).
For the parallel-magnetic-field case in
Fig.~\ref{fig:width}(b), 
Mechanism II is even more important and dominates the spin
  relaxation for $a>6$\ nm and $>2.5$\ nm when $B=5$\ T and $0.5$\ T,
  respectively. This is quite different from the cubic materials with narrower 
band
gap and larger SOC such as GaAs, in which the spin
relaxation due to Mechanism I
is usually 2-3 orders of magnitude stronger than that due to 
Mechanism II. But thanks to the small SOC, the spin relaxation caused
by the nuclei plays much more important role in GaN QD.  
It is also worth noting that the hyperfine interaction and the SOC
 can also cause spin dephasing. Previous studies on GaAs QD have shown
 that the hyperfine interaction usually dominates the spin dephasing
 at low temperature.\cite{Jiang,Johnson} It is expected that the spin
 dephasing in GaN QD is also dominated by the hyperfine interaction
 due to the very small SOC
  in this material. Our numerical results using the approach in
  Ref.~\onlinecite{Jiang} show that this is indeed true, {\em e.g.},
 for QD of $a=5$~nm and $d_{0}=10$~nm, $T_2$ induced by the hyperfine
 interaction is about five orders of magnitude shorter than that
 induced by the SOC under parallel magnetic field of $0.5$~T when
  $T=0$\ K. As we are interested in the difference between GaN and
 GaAs QDs,  we will not further address the spin dephasing
  in the paper. 
  
\begin{figure}[bth]
  \centering
  \includegraphics[width=1\columnwidth]{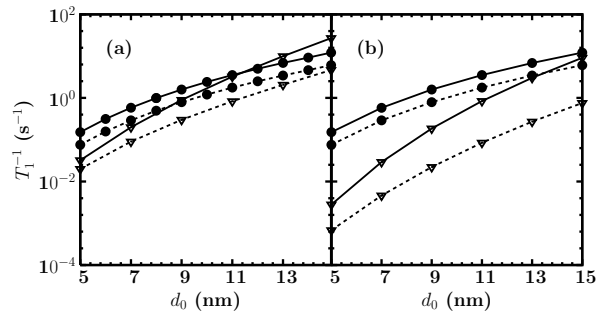}
  \caption{Spin relaxation {\em vs.} the QD
  diameter $d_{0}$ in the presence of 
 (a) perpendicular and (b) parallel magnetic fields at two well
  widths: $a=5$\ nm (solid curves) and $a=10$\ nm (dotted
  curves). In the calculation,
  $B=0.5$\ T. Curves with $\bigtriangledown$---$T_{1}^{-1}$ induced by
  the electron-phonon scattering in conjunction with the SOC; Curves
  with $\bullet$---$T_{1}^{-1}$ induced by the second-order process of the
  hyperfine interaction combined with the electron-phonon scattering.}
\label{fig:diameter}
\end{figure}

In Fig.~\ref{fig:diameter} the QD diameter dependence of the
spin relaxation is presented under the
magnetic field perpendicular (a) and parallel (b) to the well
plane. Both relaxation rates increase with the increase of dot
size but with different speeds: 
$\Gamma_{1} \propto d_{0}^{8}$ and $\Gamma_{2}\propto d_{0}^{4}$. 
As a result, Mechanism I 
becomes more important
as the size of QD grows. One can see from
Fig.~\ref{fig:diameter} that, 
under the low magnetic field ($B=0.5$~T) we show here, 
Mechanism II 
plays a very important role, or even dominates
the spin relaxation for all QD whose diameter is smaller than
$11$~nm. 

\begin{figure}[bth]
  \centering
  \includegraphics[width=1\columnwidth]{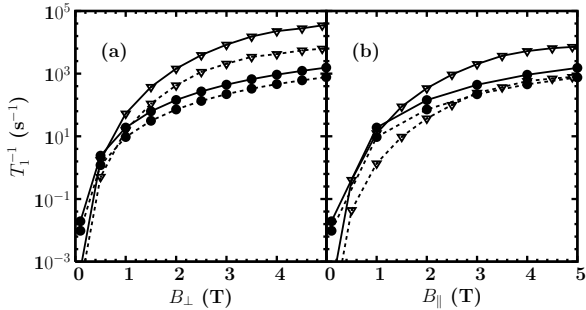}
  \caption{Spin relaxation {\em vs.} (a) perpendicular and
    (b) parallel magnetic field at two well widths:
    $a=5$\ nm (solid curves) and $a=10$\ nm (dotted curves). In the
    calculation, $d_{0}=10$\ nm. Curves with
    $\bigtriangledown$---$T_{1}^{-1}$ induced by
    the electron-phonon scattering in conjunction with the SOC; Curves
    with $\bullet$---$T_{1}^{-1}$ induced by the second-order process of the
    hyperfine interaction combined with the electron-phonon
    scattering. }
\label{fig:mag}
\end{figure}

In Fig.~\ref{fig:mag}(a) and (b) the spin relaxation rates induced by
the two mechanisms are plotted 
as functions of the perpendicular and parallel magnetic fields
respectively. In each figure, the results are shown for both narrow 
well ($a=5$\ nm) and relatively wide well ($a=10$\
nm). It is noticed that the effect of each mechanism increases with
the magnetic field as predicated by Eqs.~(\ref{gamma1}) and
(\ref{gamma2}). Then we pay attention to the relative importance
of the two mechanisms. When the magnetic field is along the
$z$-direction, it is seen from Fig.~\ref{fig:mag} that 
 Mechanism I is dominant when large vertical magnetic
  ($B>0.5$\ T) is applied. However, when the magnetic is along
  $x$-axis, for small well width ($a=5$~nm), Mechanism I
is dominant for large magnetic field. 
For wider quantum well ($a=10$\ nm), Mechanism II dominates the
  spin relaxation when $B<2.5$\ T and is comparable to Mechanism I for
  larger magnetic field.
 
\begin{figure}[bth]
\centering
  \includegraphics[width=0.6\columnwidth]{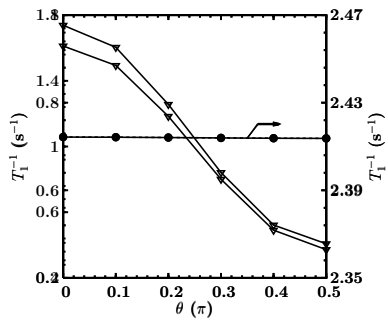}
  \caption{ Spin relaxations {\em vs.} $\theta$.
      In the calculation, $a=5$\ nm, $B=0.5$\ T and $d_{0}=10$\ nm.
      Curve with $\bigtriangledown$---$T_{1}^{-1}$ induced by
      the electron-phonon scattering in conjunction with the SOC; Curve
      with $\bullet$---$T_{1}^{-1}$ induced by the second-order process of the
      hyperfine interaction combined with the electron-phonon
      scattering. 
      Note the scale of the spin relaxation induced by
      the second mechanism is  at the right hand side of the frame.} 
  \label{fig:mag_angle}
\end{figure}

We then turn to study how the direction of the applied magnetic field
changes the spin relaxation. In Fig.~\ref{fig:mag_angle}, we show the
spin relaxation rates as functions of the angle $\theta$ between
${\mathbf B}$ and the $z$-axis for a fixed magnetic field amplitude.  
It is seen that these two mechanisms depend on the direction 
of the magnetic field quite differently. The spin relaxation induced
by Mechanism I
has a maximum 
when the magnetic field is along the $z$-direction. With the
increase of $\theta$, it decreases gradually and reaches the minimum
when ${\mathbf B}$ is in the $x$-$y$ plane. 
On the other hand, the spin
relaxation induced by Mechanism II  
almost keeps unchanged
with $\theta$. This can be understood from the
perturbation result. As we 
can see from Eqs.~(\ref{gamma1}) and (\ref{gamma2}) that the relaxation
rate $\Gamma_1$ contains the term of $(1+\cos^{2} \theta)$, which has the 
largest value for $\theta=0$ and the smallest value for
$\theta=\pi/2$ for the condition we considered, while $\Gamma_2$ is
almost independent of $\theta$. 
Overall, the changes of the spin relaxation rates in GaN QD are mild 
when the direction of the magnetic field changes for both mechanisms. 
This is quite different from  
that in GaAs QD, where the spin relaxation induced by Mechanism I
with the perpendicular magnetic field can be several
orders of magnitude larger than that with the parallel magnetic
field.\cite{Destefani,Jiang} 
This is because in GaAs material, the 
SOC is usually comparable or even larger than Zeeman splitting 
and therefore the magnetic field direction
changes the eigen energy remarkably. 
Consequently, the 
difference between the maximum and the minimum of the spin relaxation
rates induced by Mechanism I 
can be several orders of magnitude different
when the direction of the magnetic field changes.
However, due to the small SOC in GaN, the 
energy difference between the lowest two eigen states is determined by 
the Zeeman splitting and therefore the change of the spin
relaxation with the magnetic field direction is much milder. 

\begin{figure}[bth]
  \centering
  \includegraphics[width=1\columnwidth]{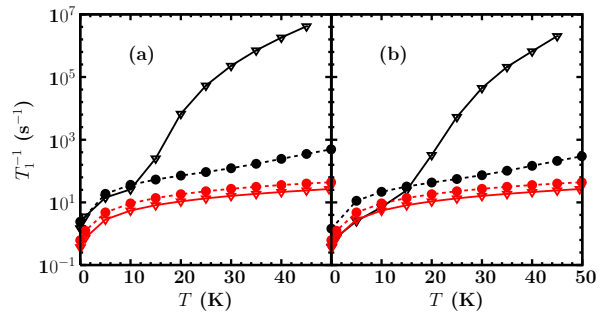}
  \caption{(Color online) Spin relaxation {\em vs.}  temperature
    $T$ in the presence of (a) perpendicular and (b) parallel magnetic fields.
 In the calculation, $a=5$\ nm, $d_{0}=10$\ nm and
    $B=0.5$\ T. Black (dark) curves---exact diagonalization results;
    Red (light) curves---perturbation
    results. 
    Curves with $\bigtriangledown$---$T_{1}^{-1}$ induced by
    the electron-phonon scattering in conjunction with the SOC; curves
    with $\bullet$---$T_{1}^{-1}$ induced by the second-order process of the
    hyperfine interaction combined with the electron-phonon scattering. }
\label{fig:temperature}
\end{figure}

We further investigate how the spin relaxation changes with the
temperature. The results are shown in 
Fig.~\ref{fig:temperature}. One can see that spin relaxations induced
by the two mechanisms both increase with 
the temperature. For low temperature regime, the relative
importance of each mechanism remains unchanged. That is, 
Mechanism I 
is more important when the magnetic field is perpendicular
to the well, while Mechanism II 
usually plays more important
role for the parallel magnetic field. Both are
approximately proportional to $[2n_{{\mathbf q}}(T) +1]$ which is
consistent with the perturbative results, 
{i.e.}, Eqs.~(\ref{phonon}) and
(\ref{second-order}). 
However, when the temperature rises high enough ($T>10$~K), 
the spin relaxation induced by Mechanism I 
increases much quicker than Mechanism II.
For both parallel and perpendicular magnetic
fields, Mechanism II
dominates the spin
relaxation for low temperature while 
Mechanism I 
has larger contribution for high temperature.  
In order to understand the different
temperature dependences of relaxations, we also show the
spin relaxation rates from perturbation calculation in the same
figure. It is seen that the perturbation result and the exact 
diagonalization result
of the spin relaxation due to 
Mechanism II 
agree with each other qualitatively
in the temperature regime we study. However, the
spin relaxation of exact diagonalization due to 
Mechanism I 
departs from the
perturbation prediction in the high temperature regime. This indicates
that the perturbation method is no longer reliable for 
Mechanism I 
and should not be used to obtain the spin
relaxation rate under high temperature. This 
is understandable, because at low temperature regime, the electron
distribution at the high levels is negligible and only the lowest two
Zeeman splitting levels are involved due to the small SOC. Therefore
perturbative method is accurate enough to study the spin relaxation
caused by Mechanism I.
With the increase of the temperature,
electron can occupy higher energy levels with larger SOC. As a
result, the perturbation method is no longer adequate to study the
transition rates induced by Mechanism I. 

\section{Conclusion}
\label{sec:conclusion}

In conclusion, we have investigated the spin relaxation
time $T_{1}$ in cubic GaN QD under different conditions by the
perturbation and exact diagonalization approaches. Two leading
spin relaxation mechanisms, {i.e.}, the
electron-phonon scattering in conjunction with the SOC and the
second-order process of the hyperfine interaction
combined with the electron-phonon scattering, are considered. 
We systematically study how the
spin relaxations induced by the two mechanisms change with the well
width $a$, magnetic field $B$ and quantum dot diameter $d_{0}$.
Our results show that, 
the ratio of these two spin relaxation rates is proportional to 
$a^{-3}B^2 d_0^4$ in the low temperature regime 
when the quantum well constraint is strong enough. Due
to the small SOC, the spin relaxation caused by the
second-order process of the hyperfine interaction combined with the
electron-phonon scattering plays much more important role in GaN
material. 
Only when the well width $a$ is small enough and/or the magnetic
field $B$ and QD diameter $d_{0}$ are large enough, the
electron-phonon scattering in conjunction with the SOC may
dominate. Furthermore, how the direction of the applied magnetic
field changes the spin relaxation are investigated. The spin
relaxation induced by the electron-phonon scattering in conjunction
with the SOC has a maximum when the magnetic field is along the
$z$-direction and reaches the minimum when the magnetic field is in the
$x$-$y$ plane. Nevertheless, the spin relaxation
 induced by the second-order process of the
hyperfine interaction combined with the electron-phonon scattering
keeps almost unchanged with the magnetic field direction. We also
discuss the temperature dependence of the spin
relaxation due to the two mechanisms.  
At high temperatures, the spin
relaxation induced by the electron-phonon scattering in conjunction
with the SOC is always dominant. 

\section{Acknowledgement}

This work was supported by the Natural Science Foundation of China
under Grants No.\ 10725417 and No.\ 10804103, the National Basic
Research Program of China under Grant No.\ 2006CB922005 and the
Innovation Project of Chinese Academy of Sciences. One of the authors (MWW) 
would like to thank X. Marie for valuable discussions.

\end{document}